\documentstyle[floats,epsf,aps]{revtex}
\epsfclipon

\begin{document}
\draft

\title{Conductance fluctuations in mesoscopic
normal-metal/superconductor samples}

\twocolumn[\hsize\textwidth\columnwidth\hsize\csname @twocolumnfalse\endcsname

\author{Klaus Hecker$^1$, Helmut Hegger$^1$, Alexander Altland$^{2,3}$,
and Klaus Fiegle$^4$}

\address{$^1$ II.\,Physikalisches Institut, Universit\"at zu K\"oln,
D-50937 K\"oln, Germany\\
$^2$ Institut f\"ur Theoretische Physik, Universit\"at zu K\"oln,
D-50937 K\"oln, Germany\\
$^3$ Cavendish Laboratory, Madingley Road, Cambridge CB3 0HE, UK\\
$^4$ I.\,Physikalisches Institut, Universit\"at zu K\"oln,
D-50937 K\"oln, Germany\\}

\date{\today{}}

\maketitle

\begin{abstract}

We study the magnetoconductance fluctuations of mesoscopic
normal-metal/superconductor (NS) samples consisting of a gold-wire in
contact with a niobium film. The magnetic field strength is varied over a
wide range, including values that are larger than the upper critical
field $B_{c2}$ of niobium. In agreement with recent theoretical
predictions we find that in the NS sample the rms of the conductance
fluctuations (CF) is by a factor of $2.8 \pm 0.4$ larger than in the high
field regime where the entire system is driven normal conducting. Further
characteristics of the CF are discussed.  

\end{abstract}



]

At low temperatures disordered metals smaller than the phase coherence
length $L_\varphi$ (mesoscopic systems) exhibit a host of
quantum fluctuation phenomena like e.g. universal fluctuations of the
electrical conductance\cite{lst85,lsf87,wwe92}. Recently it has become
clear that additional mechanisms of quantum coherence arise when a
mesoscopic normal metal sample (N) is brought in contact with a
superconductor (S)
\cite{azi97,bbe95b,kkg91,pad95,hkw96,cgm96,gpb96,ccg96,teb92b,mbj93,bbe95,nst96,val96}.
These processes, which are caused by the interference of electrons and
Andreev reflected holes \cite{and64}, manifest themselves in the
emergence of additional universality classes \cite{azi97} and an altered
fluctuation behaviour of various mesoscopic observables. In particular
the fluctuations of the magneto-conductance are still universal of
order $e^2/h$ but tend to exceed the normal-metal fluctuations by
numerical factors of order unity. These enhancement factors are a
consequence of (i) the fact that {\it two} elementary charges are
transferred per Andreev reflection \cite{azi97,bbe95b,teb92b,mbj93} and
(ii) the presence of diffusionlike modes which are absent in the pure
N-case. To be specific, it has been found that in the presence of an
external magnetic field the rms amplitude of the conductance
fluctuations $ {\rm rms}(G_{\rm NS})$ exceeds 
its normal metal $ {\rm rms}(G_{\rm N})$ value by a factor of
$2\sqrt{2}$. This result holds true in the presence \cite{azi97} or
absence \cite{azi97,bbe95b} of spin-orbit scattering.

\begin{figure}[t]
  \begin{center}
    \leavevmode
    \epsfxsize=7cm
    \par
    \epsffile{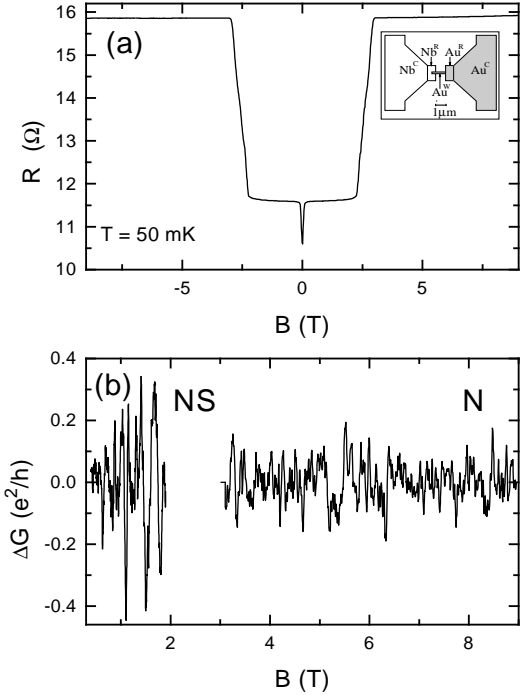}
    \par
  \end{center}
\caption{(a) Magnetoresistance of sample 1. Three regimes can be
identified (i) For small magnetic fields ($B_1 \simeq 0.1$ T) a
proximity effect induced resistance minimum is observed. (ii) In the
field range $B \simeq 0.4-2$ T the resistance is approximately
constant. (iii) For $B>3.1$~T the whole sample is driven normal
conducting. The inset shows the sample geometry: A mesoscopic Au wire
(Au$^{\rm W}$, length $L \simeq 1 \; \mu$m, width $W \simeq 180$ nm)
is connected to a Nb contact (Nb$^{\rm C}$+Nb$^{\rm R}$) and an
Au contact (Au$^{\rm R}$+Au$^{\rm C}$) (mesoscopic two-probe
arrangement). (b) Conductance fluctuations of sample 1 in the NS-case
(0.4 T$<B<$1.9 T) ${\rm rms}(G_{\rm NS})= 0.16 \pm 0.02 \; e^2/h$ and
the N-case (3.1 T$<B<$9 T) ${\rm rms}(G_{\rm N})= 0.058 \pm 0.003 \;
e^2/h$.}
\label{figRB}
\end{figure}

The fluctuation characteristics of a NS sample are strongly affected
by magnetic fields. At least three different regimes with
qualitatively different behaviour can be identified [cf.
Fig.\ref{figRB}(a)]: (i) The presence of the proximity effect gives
rise to a resistance minimum for small magnetic fields. In our samples
the proximity effect is supressed by fields of order $B_1 \sim 0.1$ T
\cite{foot1}.  (ii) For intermediate fields larger than $B_1$ but
smaller than the upper critical field of the superconductor $B_{c2}$, the
proximity effect is supressed but the process of Andreev reflection is
still active. (iii) For fields larger than $B_{c2}$ the
superconductivity is globally destroyed.

Most experiments of other groups were performed in regime
(i)\cite{pad95,hkw96,cgm96,gpb96,ccg96}.  In contrast, we have studied
the magneto-CF in the regimes (ii) and (iii).  We are thus in a
position to compare the fluctuation behaviour of {\it one and the
same sample} in both regimes NS and pure N. As a result, we find that
the CF of the NS system are enhanced as compared to those of the N
system. The ratio $ {\rm rms}(G_{\rm NS})/{\rm rms}(G_{\rm N})$ turns
out to be in good agreement with the theoretical prediction.

The inset of figure \ref{figRB}(a) shows the sample layout. The sample
consists of a Nb contact, a mesoscopic Au wire and a second Au
contact.  This layout corresponds to a mesoscopic two-probe
arrangement \cite{hhs94}.  Far outside the coherence volume the two
contacts are split-up to perform a macroscopic four-point measurement.
The samples were prepared in two successive steps using electron-beam
lithography and lift-off technique \cite{fdj97}.  In a first step the
Nb contact was prepared by in situ deposition of a 30 nm Nb and a 10
nm Au layer, which prevents oxidation of the Nb. In a second process
the Au wire of length $L \simeq 1 \; \mu$m (width $W\simeq 180$ nm,
thickness $t\simeq 30$ nm) and the Au contact was produced. Both the
Nb and Au layers were dc magnetron sputtered. We report on
measurements of two samples which had a normal state resistance
$R_{\rm N} \sim 15 \; \Omega $ at low temperatures (see Table
\ref{tab1}).  From the residual resistance ratio and from the sheet
resistance of wide Nb and Au films of the same thickness we derive
elastic mean free paths of $\ell_{Au} \simeq 32$ nm, $\ell_{Nb} \simeq
5$ nm and diffusion constants of $ D_{Au} \simeq 150 \; {\rm cm^2/s}$,
$ D_{Nb} \simeq 23 \; {\rm cm^2/s}$. The Thouless energy $E_{Th}= h
D_{Au} / L^2_\varphi $ for our systems is $E_{Th} \simeq 100 \; \mu$eV
\cite{tho77}.  The critical temperature of the Nb is $T_c \simeq 8$ K,
for Nb films with and without a Au layer, and $B_{c2}(T = 100 \; {\rm
  mK}) \simeq 2.5$ T. The measurements were performed in a
$^3$He--$^4$He-dilution refrigerator at temperatures down to $T_{min}
\simeq 45 $ mK \cite{hhs94}. The conductance $G(B)$ was measured by
means of a standard lock-in technique for magnetic fields up to $B=9$
T. We used small measurement currents $I_{ac}=1-3 \; \mu$A at
temperatures below 1K.  Larger currents were used at $T >1$ K always
under the condition $V_{ac} < k_BT/e$.

\begin{table}[h]
\caption{$R_{\rm N}$: normal state resistance at $T = 50$ mK and $B = 4$ T,
$R_{\rm NS}$: resistance in the intermediate regime (ii) at $T = 50$
mK and $B = 1$ T {\protect (see Fig.\ref{figRB})}.}
\begin{tabular}{lcc}
\narrowtext
& sample 1 & sample 2 \\
\tableline
$R_{\rm N}$ ($\Omega$)               & 15.87             & 14.34               \\
$R_{\rm NS}$ ($\Omega$)            & 11.60             & 9.72             \\
${\rm rms}(G_{\rm NS}) $ ($e^2/h$) & 0.16 $\pm 0.02 $  & 0.14 $ \pm 0.02$ \\
${\rm rms}(G_{\rm N}) $ ($e^2/h$)  & 0.058 $\pm 0.003$ & 0.050 $ \pm
0.004$ \\
${\rm rms}(G_{\rm NS})/{\rm rms}(G_{\rm N}) $ & 2.8 $\pm 0.4 $ & 2.8 $\pm
 0.4 $ \\
\end{tabular}
\narrowtext
\label{tab1}
\end{table}

We now turn to the discussion of our results.  Figure \ref{figRB}(b)
shows the conductance fluctuations for both the NS- and the N-case as
a function of the magnetic field. Since the magnetoconductance is
measured in a mesoscopic two-probe configuration, its fluctuations are
symmetric with respect to a reversal of the magnetic field
\cite{hhs94,hhh96}. For this reason only the part $B>0$ T of the
magnetofingerprints $\Delta G (B)$ is shown. The NS-CF at low magnetic
fields are clearly larger than in the N-case ($B \geq 3.1$ T).

\begin{figure}
  \begin{center}
    \leavevmode
    \epsfxsize=7cm
    \par
    \epsffile{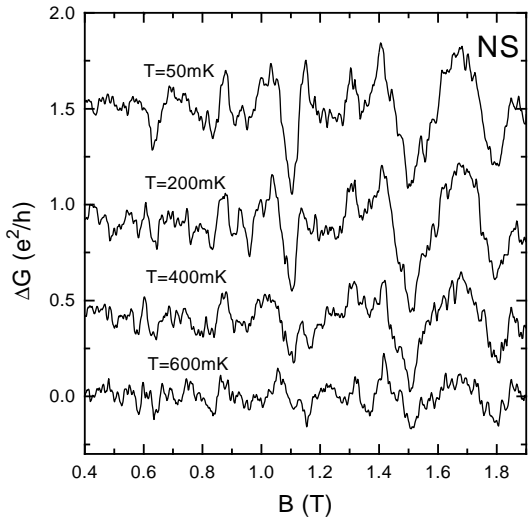}
    \par
  \end{center}
\caption{NS-conductance fluctuations of sample 1 for different
temperatures. For temperatures $T \geq 300$ mK the CF are reduced
(cf. Fig.\ref{figRMS}). Traces are shifted for clarity.}
\label{figNSCF}
\end{figure}

\begin{figure}
  \begin{center}
    \leavevmode
    \epsfxsize=6cm
    \par
    \epsffile{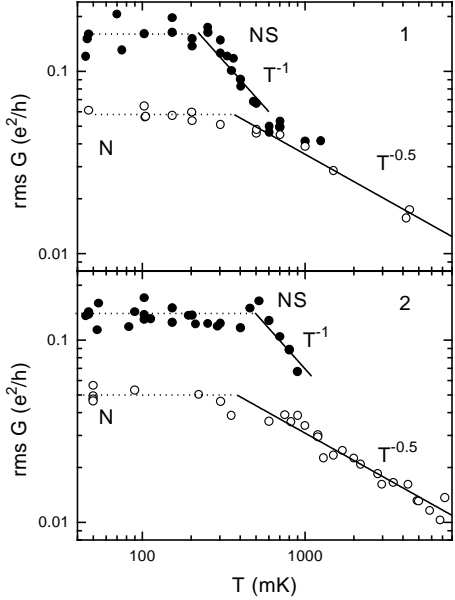}
    \par
  \end{center}
\caption{rms amplitudes of the NS-CF ($\bullet$) and the N-CF
($\circ$) for sample 1 (upper Fig.) and sample 2 (lower Fig.). The
dotted lines show the low temperature saturation values (cf.
tab.\ref{tab1}). The reduction of ${\rm rms}[G_{\rm NS}(T)]$ starts at
about 300-500 mK. The N-CF sets are reduced for $T \geq 400$ mK.  The
solid lines correspond to $T^{-1}$ (NS) and $T^{-0.5}$ (N) power
laws.}
\label{figRMS}
\end{figure}

We obtain values of ${\rm rms}(G_{\rm NS1})= 0.16 \pm 0.02 \; e^2/h$
and ${\rm rms}(G_{\rm N1})= 0.058 \pm 0.003 \; e^2/h$ for sample 1
(see table \ref{tab1} and
Figs. \ref{figRB}-\ref{figRMS}). For sample 2 the
rms values are slightly smaller: ${\rm rms}(G_{\rm NS2})= 0.14 \pm
0.02 \; e^2/h$ and ${\rm rms}(G_{\rm N2})= 0.050 \pm 0.004 \; e^2/h$
(see table \ref{tab1} and Fig.\ref{figRMS}). Since the
magnetic-field range for the calculation of the NS-CF is much smaller
($\Delta B_{\rm NS} = 1.5$ T) than the range of the N-CF ($\Delta
B_{\rm N} = 5.9$ T) the uncertainty for the ${\rm rms}(G_{\rm NS})$
values is substantially larger.

We note that the measured values of the CF are much smaller than the
results obtained in zero temperature theoretical calculations. To
understand better the origin of this discrepancy, viz. the combined
effect of finite temperatures and dephasing effects, let us briefly
recall a few known theoretical predictions about the CF of diffusive
mesoscopic systems. As has been shown diagrammatically
(cf.\cite{lsf87} and references therein) the CF of
quasi--one--dimensional wires in the presence of both spin-orbit
interactions and a magnetic field are given by

\begin{eqnarray}
&&\left.\begin{array}{l}
{\rm rms}(G_{\rm N})^2\\
{\rm rms}(G_{\rm NS})^2
\end{array}\right\} = \frac{6e^4}{h^2}\left\langle
\sum_q \left(
\frac{hD/L_{\rm eff}^2}{h Dq^2 + h/\tau_\varphi + i\epsilon} \right)^2
\right\rangle_{\epsilon}\times\nonumber\\
&&\hspace{3.0cm}\times\left\{
\begin{array}{l}
1\\
8
\end{array}\right..
\label{rmsformel}
\end{eqnarray}

Here $\langle \dots \rangle_{\epsilon} := \frac{1}{2k_BT}
\int_{-k_BT}^{k_BT} d\epsilon (\dots)$ stands for a temperature
induced energy averaging procedure\cite{lsf87}, $\sum_q$ denotes a
summation over quantized momenta $q=\pi/L_{\rm eff}, 2\pi/L_{\rm eff},
\dots$ and it has been assumed that the wire's cross section
$L_{\perp}\ll L_{\rm eff}$ (i.e.  $ h D/L_{\perp}^2$ is by far larger
than any relevant energy scale in the problem which means that the
wire can be regarded as 'quasi'--one--dimensional\cite{lsf87}).  The
phenomenological parameter $\tau_\varphi$
($L_\varphi=\sqrt{D\tau_\varphi}$) accounts for the various dephasing
mechanisms which lead to a destruction of the conductance
fluctutions. Whereas the formula for the N-case can be found at
various places in the literature (cf. e.g. \cite{akl86} for a
reference with correct prefactors) the NS-formula is less standard and
deserves a few comments. The relative factor of eight results from (a)
the fact that two elementary charges are driven through the system
whenever an Andreev reflection occurs ($e\rightarrow 2e$) and (b) the
presence of twice as many diffusionlike modes as in the
N-regime. These additional modes are in many respects similar to the
standard diffusive modes (cf. the discussion in Ref.\cite{azi97}),
that is their presence manifests in an additional factor of two
(rather then in a structurally altered formula). We emphasize that in
a normal metal, the scale $L_{\rm eff}$ appearing in (\ref{rmsformel})
represents the {\it effective} length of the sample, i.e., the length
of the region where most of the voltage drops. In a NS-sample,
however, $L_{\rm eff}$ is {\it twice} that length. This is a
consequence of the fact that both the incoming electrons and the
outgoing holes have to traverse the system
diffusively\cite{bbe95b,teb92b}.

We next discuss the absolute magnitude of the CF.  Equation (\ref{rmsformel})
contains most of the information about conductance fluctuations we need to
know. In the limit of low temperatures and vanishing inelastic interactions
($T,\tau_\varphi^{-1} \rightarrow 0$), Eq.(\ref{rmsformel}) yields the
universal values ${\rm rms}(G_{\rm N}^{theo}) = 0.258 \; e^2/h$ and ${\rm
rms}(G_{\rm NS}^{theo}) = 0.729 \; e^2/h$.  In our experiment, however, the
low temperature regime is characterized by the inequality $L_T>L_{\rm eff}\gg
L_\varphi$, where $L_\varphi=( D \tau_\varphi )^{1/2}$ and $L_T= (h D /
k_B T)^{1/2}$ are the dephasing and thermal length, respectively. Under
these circumstances the conductance fluctuations are reduced by a factor
$\sim (L_\varphi/L_{\rm eff})^{(4-d)/2}$ below their universal, length
independent value ($d$: dimensionality of the system). In order to
interpret the experimental data we therefore carefully have to estimate the
scale $L_{\rm eff}$, i.e., the size of the sample region that contributes most
to the resistance (or equivalently to the conductance fluctuations). 

It is evident from the sample geometry shown in the inset of Fig.
\ref{figRB}(a), that the main part of the voltage drop occurs in the
narrow wire Au$^{\rm W}$. The resistance of the wide contacts
(Nb$^{\rm C}$+Nb$^{\rm R}$ and Au$^{\rm R}$+Au$^{\rm C}$ in
Fig.\ref{figRB}(a)) is about $1.2 R_\Box$ each ($R_\Box$: sheet
resistance). The main voltage drop of a contact is located at the
small rectangles Nb$^{\rm R}$ and Au$^{\rm R}$ corresponding to a
resistance of about $0.5 R_\Box$. The sheet resistances of the Nb and
the Au differ by a factor of about four ($R_\Box^{Nb} \simeq 4 \;
\Omega$, $R_\Box^{Au} \simeq 0.9 \; \Omega$).  Thus the contribution
of the Au rectangle Au$^{\rm R}$ is about $ 0.5 \; \Omega$ which is
less than 5 \% of the total resistance and neglected in the following.
Hence, in the NS-case the voltage drop can be attributed to the wire
Au$^{\rm W}$. We obtain an effective length of $L_{\rm eff}^{NS}
\simeq 2L \simeq 2 \,\mu$m.  In the case where the whole sample is
normal conducting the situation is different.  The Nb rectangle
Nb$^{\rm R}$ has a resistance of about $2 \; \Omega$ corresponding to
13 \% of the total resistance. The region contributing to the voltage
drop is thus larger in the N-case. Here, $L_{\rm eff}$ is given by the
regions Au$^{\rm W}$ {\em and} Nb$^{\rm R}$ resulting in $L_{\rm
eff}^{\rm N} \simeq 1.8 \mu$m.  Of course this is only a rough
estimate for the effective lengths of the samples. The main conclusion
to be drawn is that the effective lengths $L_{\rm eff}$ are {\it
roughly the same in both the N-case and the NS-case}. To check the
validity of the above consideration we have calculated the phase
coherence length of the system via $L_\varphi = L_{\rm eff} [{\rm
rms}(G_{exp}) /{\rm rms}(G_{theo})]^{2/3}$ using the previously
estimated values of $L_{\rm eff}$.  The dephasing length $L_\varphi$
is expected to be roughly independent of the strength of the magnetic
field (as long as the latter is sufficiently strong to break time
reversal invariance). Indeed we obtain almost constant values $
L_\varphi \simeq 660 \pm 60 $ nm for the different regimes and
samples\cite{foot2}, thereby supporting the correctness of our
estimates for $L_{\rm eff}$. Furthermore the value for $L_\varphi$ is
in good agreement with previous studies \cite{hhs94,hhh96}.

Since $L_{\rm eff}^{\rm NS} \simeq L_{\rm eff}^{\rm N} $, we can
directly compare the respective rms values of the CF. We obtain a
ratio of

\begin{equation}
\frac{{\rm rms}(G_{\rm NS})}{{\rm rms}(G_{\rm N})} = 2.8 \pm 0.4  \label{2}
\end{equation}

for both samples (see table \ref{tab1}).  This ratio is in agreement
with the theoretical value of $2\sqrt{2}$ \cite{azi97,bbe95b}. Earlier
theoretical work predicted an enhancement by only a factor two
\cite{teb92b}. In contrast, the experiment confirms the recent
results of Altland and Zirnbauer \cite{azi97} and Brouwer and
Beenakker \cite{bbe95b}. Equation (\ref{2}) represents the central result
of our experiment. In the next few paragraphs we report on further
characteristics of the NS fluctuations, in particular on their
temperature and voltage dependence.

Figure \ref{figNSCF} shows NS-magnetofingerprints in the temperature range
beween 50~mK and 600~mK. Up to $T \simeq 300$ mK we observe the low
temperature saturation value of ${\rm rms}(G_{\rm NS})= 0.16 \pm 0.02 \;
e^2/h$, as shown in Fig.\ref{figRMS}. For higher temperatures the CF are
supressed and follow a $T^{-1\pm 0.1}$ law. A similar behavior is observed
for sample 2 (cf. Fig. \ref{figRMS}). Here the reduction starts at a
somewhat higher temperature $T\simeq 400$ mK.  The N-CF were investigated
for temperatures between 45 mK and 7 K (cf. Fig. \ref{figRMS}). The
reduction of the N-CF starts at $T \geq 400$ mK and follows a $T^{-0.50\pm
0.05}$ law for both samples. 

In general, CF are supressed as soon as either $ L_\varphi(T) $ or
$L_T$ become smaller than $ L_{\rm eff} $ (cf. Eq.\ref{rmsformel}). For
higher temperatures they show a weak power law behavior\cite{lsf87}
${\rm rms}[G(T)] \sim T^{-\alpha}$. The exponent depends on various
system characteristics such as the ratio between the different length
scales, and its dimensionality. However, we do not have a compelling
argument for why $\alpha $ should depend on the presence or absence of
superconductivity. In other words, a conclusive explanation of the
different power-law behaviours observed in the experiment is
lacking. Nevertheless, we would like to speculate on a mechanism that
{\it might} be responsible for this effect: We observe that the
rms amplitudes of the NS-CF and the N-CF saturate at low temperatures
although $L_{\rm eff}> L_\varphi $ in our samples.  We therefore conclude
that $ L_\varphi $ must be almost temperature independent at low
temperatures. If the temperature is raised, $L_T$ and $L_\varphi$
decrease and thereby the CF. The point now is that the decay rate of
the CF sensitively depends on the effective dimensionality of the
sample -- the lower the dimensionality, the higher the decay rate (cf.
Ref. \cite{lsf87} for details). In regime (ii) our systems are clearly
quasi--one--dimensional. However in (iii) the situation is different
since the effectively two--dimensional wide Nb electrode contributes to
the CF.  This might give rise to a less pronounced temperature
dependence.  Nonetheless it seems to be unlikely that a dimensional
crossover of this kind can account for the whole effect.

Let us finally comment on the dependence of the CF on the measuring
voltage. Voltages $V$ exceeding the value $V_{Th}=E_{Th}/e$ break the
symmetry between particles and holes thereby leading to a destruction
of the above mentioned additional diffusive modes. In our samples
$V_{Th}\simeq 100 \; \mu$V\cite{azi97,bbe95b,mbj93,bbe95}.  Indeed we
do not observe any influence of $V_{ac}$ on size or temperature
dependence of the NS-CF up to $V_{ac} = 35 \; \mu$V, which is of the
same order as $V_{Th}$. For higher voltages ${\rm rms}(G_{\rm NS})$ is
reduced.

To summarize, we have measured the magnetoconductance fluctuations of
mesoscopic Au/Nb systems.  By changing the magnetic field strength we
induced a cross\-over from a normal-/superconducting to a purely
normalconducting state.  We found a relative enhancement factor ${\rm
rms}(G_{\rm NS})/{\rm rms}(G_{\rm N}) = 2.8 \pm 0.4 $ in good
agreement with the theoretical prediction ${\rm rms}(G_{\rm
NS}^{theo})/{\rm rms}(G_{\rm N}^{theo}) = 2\sqrt{2} $
\cite{azi97}. For large temperatures the NS-CF behave like $ {\rm
rms}[G_{\rm NS}(T)] \sim 1/T $ thereby showing a stronger temperature
dependence than the N-CF.

We thank R.~Gross, K.~Jacobs, D.E.~Khmelnitskii, H.~Micklitz,
A.~Nowack, and M.~Zirnbauer for useful discussions. This work was
supported by the Deutsche Forschungsgemeinschaft through SFB 341 and
SFB 301 and by BMBF, grant 052KV134-(6) and 053KU234-(0).


\begin{references}
\bibitem{foot1} In zero magnetic field a reentrance of metallic
conductivity occurs at low temperatures and voltages \cite{ccg96}. We
observe a resistance minimum at $T \simeq 1$ K in $R(T)$ and at
$V_{dc} \simeq 100 \; \mu$V in IV-characteristics.
\bibitem{foot2} As we do not observe a weak-localization contribution to
the conductance the exact value of $L_\varphi$ cannot be evaluated in our
samples.
\end{references}
\end{document}